\documentclass[useAMS,usenatbib,usegraphicx]{mnras}
\usepackage{graphicx,amssymb,amsmath,times,verbatim}
\usepackage[normalem]{ulem}

\def\lta{\mathrel{\spose{\lower 3pt\hbox{$\mathchar"218$}}
     \raise 2.0pt\hbox{$\mathchar"13C$}}}
\def\gta{\mathrel{\spose{\lower 3pt\hbox{$\mathchar"218$}}
     \raise 2.0pt\hbox{$\mathchar"13E$}}}

\def\mathnew{\mathsurround=0pt}

\def\simov#1#2{\lower .5pt\vbox{\baselineskip0pt \lineskip-.5pt
\ialign{$\mathnew#1\hfil##\hfil$\crcr#2\crcr\sim\crcr}}}

\title[Multi-wavelength features of OJ 287 during 2017]{The ever-surprising blazar OJ 287:
multi-wavelength study and appearance of a new component in X-rays}

\author[Kushwaha et al.]{Pankaj Kushwaha$^{1}$\thanks{E-mail: pankaj.kushwaha@iag.usp.br},
Alok C. Gupta$^{2,3}$\thanks{E-mail: acgupta30@gmail.com}, 
Paul J. Wiita$^{4}$\thanks{Email: wiitap@tcnj.edu}, Main Pal$^{5}$,  Haritma Gaur$^{3}$,
\newauthor 
E.~M. de Gouveia Dal Pino$^{1}$, O.~M.~Kurtanidze$^{6,7,8,9}$,E.~Semkov$^{10}$, G.~Damljanovic$^{11}$,
\newauthor 
S. M. Hu$^{12}$, M.~Uemura$^{13}$, O.~Vince$^{11}$, A. Darriba$^{14,15}$, M. F. Gu$^{3}$, R.~Bachev$^{10}$, 
\newauthor
Xu Chen$^{12}$, R. Itoh$^{16}$, M. Kawabata$^{13}$, S. O. Kurtanidze$^{6}$, T. Nakaoka$^{13}$,  
\newauthor 
M. G. Nikolashvili$^{6}$, L. A. Sigua$^{6}$, A.~Strigachev$^{10}$, Z. Zhang$^{17}$  \\
\\
$^{1}$Department of Astronomy (IAG-USP), University of Sao Paulo, Sao Paulo 05508-090, Brazil \\
$^{2}$Aryabhatta Research Institute of Observational Sciences (ARIES), Manora Peak, Nainital 263002, India \\
$^{3}$Key Laboratory for Research in Galaxies and Cosmology, Shanghai Astronomical
Observatory, Chinese Academy of Sciences, \\
~~~80 Nandan Road, Shanghai 200030, China \\
$^{4}$Department of Physics, The College of New Jersey, P.O.\ Box 7718, Ewing, NJ 08628-0718, USA  \\
$^{5}$Astronomy \& Astrophysics Division, Physical Research Laboratory, Ahmedabad 380009, India \\
$^{6}$Abastumani Observatory, Mt. Kanobili, 0301 Abastumani, Georgia \\
$^{7}$Engelhardt Astronomical Observatory, Kazan Federal University, Tatarstan, Russia \\
$^{8}$Center for Astrophysics, Guangzhou University, Guangzhou 510006, China \\
$^{9}$Key Laboratory of Optical Astronomy, National Astronomical Observatories, Chinese Academy of Sciences,
Beijing 100012, China \\
$^{10}$Institute of Astronomy and National Astronomical Observatory, Bulgarian Academy of Sciences, 72 Tsarigradsko 
Shosse Blvd., 1784 Sofia, Bulgaria \\
$^{11}$Astronomical Observatory, Volgina 7, 11060 Belgrade, Serbia \\
$^{12}$Shandong Provincial Key Laboratory of Optical Astronomy and Solar-Terrestrial Environment, Institute of Space
Sciences, Shandong University, \\
~~~~Weihai 264209, China \\
$^{13}$Hiroshima Astrophysical Science Center, Hiroshima University, Kagamiyama 1-3-1, Higashi-Hiroshima 739-8526, Japan \\
$^{14}$American Association of Variable Star Observers (AAVSO), 49 Bay State Rd., Cambridge, MA 02138, USA \\
$^{15}$Group M1, Centro Astron$\acute{o}$mico de Avila, Madrid, Spain \\
$^{16}$Department of Physics, Tokyo Institute of Technology, 2-12-1 Ookayama, Meguro-ku, Tokyo 152-8551, Japan \\
$^{17}$Shanghai Astronomical Observatory, Chinese Academy of Sciences, 80 Nandan Road, Shanghai 200030, China \\
}

\begin{document}

\maketitle

\begin{abstract}
 We present a multi-wavelength spectral and temporal investigation of OJ 287
emission during its strong optical-to-X-ray activity between July 2016 - July 2017.
The daily $\gamma$-ray fluxes from \emph{Fermi}-LAT are consistent with no variability.
The strong optical-to-X-ray variability
is accompanied by a change in power-law spectral index of the X-ray spectrum from
$< 2$ to $>2$, with variations often associated with changes in optical polarization
properties. Cross-correlations between optical-to-X-ray emission during four continuous
segments show simultaneous optical-ultraviolet (UV) variations while the X-ray
and UV/optical are simultaneous only during the middle two segments. In the first
segment, the results suggest X-rays lag the optical/UV, while in the last segment X-rays
lead by $\sim$ 5-6 days. The last segment also shows a systematic trend
with variations appearing first at higher energies followed by lower energy ones. The 
LAT spectrum before the VHE activity is similar to preceding quiescent state spectrum
while it hardens during VHE activity  period and is consistent with the extrapolated
VHE spectrum during the latter. Overall, the broadband spectral energy 
distributions (SEDs) during high activity periods are a combination of a typical OJ 287 SED and an HBL
SED, and can be explained in a two-zone leptonic model, with the second zone located
at parsec scales, beyond the broad line region, being responsible for the HBL-like
spectrum. The change of polarization properties from systematic to chaotic and
back to systematic, before, during and after the VHE activity, suggest dynamic
roles for magnetic fields and turbulence.

\end{abstract}



\section{Introduction} \label{sec:intro}
Blazars are the high and rapidly variable subclass of active galactic nuclei (AGNs)
with emission dominated by non-thermal radiation emanating from relativistic jets
oriented at small angles to our line of sight. The variability activity is typically 
incoherent and complex, covers all the accessible observational domains from radio
to GeV/TeV $\gamma$-rays with temporal variation from minutes to decades, polarization
from a few to $\sim 70$\%, and spatially extended jets up to Mpc scales. The Doppler
boosting associated with bulk relativistic motions in the jets severely affects the
spectral and temporal domains, making the variability appear to be the most
energetically extreme among persistently variable astrophysical objects. Historically,
blazars have been classified empirically as BL Lacartae objects and flat spectrum radio
quasars (FSRQs) based on the properties of their optical-ultraviolet
(UV) spectra. The former is characterized by very weak emission lines or even a
featureless spectrum while the latter exhibits prominent broad emission lines 
\citep[e.g.][]{1995PASP..107..803U}.

Unlike the apparent stochastic temporal variability \citep{2017arXiv170904457G,
2016ApJ...822L..13K,2017ApJ...849..138K}, the spectral domain exhibits a characteristic
spectral energy distribution (SED) dominated by two broad humps. The first hump
peaks somewhere between the near-infrared (NIR) to UV or soft-X-ray bands, while
the second hump peaks at $\gamma$-ray energies. This characteristic has led to the
classification of blazars in terms of the location of the first hump. Thus, BL Lacs
have been named as low-energy peaked (LBL), intermediate-energy peaked (IBL) and
high-energy peaked (HBL) BL Lacs, respectively \citep{1998MNRAS.299..433F,1995ApJ...444..567P}
while FSRQs, so far, have been found to be exclusively low-energy peaked. The radiation
from the first hump  is well understood to be the synchrotron emission from relativistic
non-thermal electrons while the processes for the high energy emission are still
uncertain. Both leptonic and hadronic based processes can reproduce the high hump
fairly well. In the leptonic scenarios, the high energy hump results from inverse
Compton (IC) scattering of the surrounding  photon fields \citep[e.g.][]{2009MNRAS.397..985G};
e.g.\ for synchrotron photons themselves this is called the synchrotron-self-Compton
(SSC) process whereas for fields external to the jet, such as from the accretion disk
or broad line region this is external Comptonization (EC). The hadronic picture
invokes proton-synchrotron and/or proton-photon cascade processes \citep[e.g.][]
{2003APh....18..593M}.

 OJ 287 is a BL Lacartae object with an LBL SED, located at the redshift of $\rm 0.306$. It is one 
of the bright and highly variable sources in radio and optical bands. The source has been observed
to exhibit a very wide range of variations in all the domains and is best known for
its regular $\sim$ 12 years outbursts in optical R-band data available since 1890
\citep[and references therein] {2013A&A...559A..20H,2016ApJ...819L..37V}. This regular
feature has been used to claim it to be a binary super-massive black hole (SMBH)
system \citep[and references therein]{1988ApJ...325..628S,2016ApJ...819L..37V} with
the major outbursts arising from the impact of the secondary black hole on the
accretion disk of the primary. The model has evolved from its original inception
with observational inputs and has been successful in predicting the timing of the
regular flares quite well. In addition, many other quasi-periodicities of different
durations have been claimed for OJ 287 e.g. 40--50 days \citep{2013MNRAS.434.3122P,
2006AJ....132.1256W}, $\rm\sim$435 days \citep{2016AJ....151...54S}, and $\rm\sim$400
days \citep{2016ApJ...832...47B}.

Being bright at radio and optical energies with intriguing variability features,
OJ 287 has been followed extensively and intensively in different energy bands,
including exploration of jet features and emission in radio \citep{2017AstL...43..796M,
2013ApJS..205...15T,2012ApJ...747...63A,2011AJ....141..178M,2004ApJ...608..149T},
optical \citep{2010MNRAS.402.2087V}, UV and X-rays \citep{2017MNRAS.468..426S,
2011ApJ...729...26M}, along with its broadband emission variability and spectrum
\citep{2009PASJ...61.1011S,2011ApJ...726L..13A,2013MNRAS.433.2380K,2018MNRAS.473.1145K}.
On long terms, jet kinematic studies at radio wavelengths show complex patterns
with a sharp swing of $> 100^\circ$ in the projected jet-position-angle and jet
precession/wobbling \citep{2012ApJ...747...63A,2004ApJ...608..149T}. The optical/UV
emission has been found to lack any clear relation vis-{\`a}-vis X-ray, with optical
variability being more pronounced than the X-ray \citep{2017MNRAS.468..426S}.
Additionally, \citet{2017MNRAS.468..426S} also found a harder-when-brighter
spectral behavior in the X-ray but not in the optical band. On short terms, the
multi-wavelength emission has been found to vary almost simultaneously from NIR
to $\gamma$-ray energies \citep{2011ApJ...726L..13A, 2013MNRAS.433.2380K,2018MNRAS.473.1145K}.
Normally in BL Lacs, the high energy emission is argued to be the result of SSC.
However, using a systematic analysis during a very well observed flare of OJ 287
in 2009, \citet{2013MNRAS.433.2380K} showed that SSC cannot explain the high energy
emission \citep[see also][]{2009PASJ...61.1011S} and argued that it originated via
an EC process. This interpretation is consistent with the recent detection of new
components in NIR-optical and optical-UV with a hardening of the $\gamma$-ray spectra
\citep{2018MNRAS.473.1145K} compared to previous ones \citep{2009PASJ...61.1011S,
2013MNRAS.433.2380K}.

The latest optical outburst of OJ 287 produced by an impact on the accretion disc
was expected to occur in December 2015 in the binary SMBH model \citep[and references
therein]{2010ApJ...709..725V}. Hence, the source was being monitored and indeed the
hint of first enhancement/activity at IR-optical energies was noted around November
14, 2015 \citep[MJD 57340;][]{2016ApJ...819L..37V,2017MNRAS.465.4423G} and was
subsequently followed intensively by many research groups and observatories across
the globe in both photometry and polarization \citep{2016ApJ...819L..37V,2017MNRAS.465.4423G,
2017ApJ...835..275R,2018MNRAS.473.1145K}. The predicted disk impact flare was observed
on 5th December 2015 showing a relatively low flux polarization \citep[$<$ 10\%;][]
{2016ApJ...819L..37V}. Concurrent to this, activity across the electromagnetic
spectrum was observed, with increased activity at X-ray and $\gamma$-ray energies
as well \citep{2018MNRAS.473.1145K}. A detailed  investigation of multi-wavelength
spectral and temporal variability by \citet{2018MNRAS.473.1145K} during November
2015 -- May 2016, for the first time revealed a spectral bump in the optical-IR
and another in the optical-UV spectrum along with a hardening and shift in the peak
of $\gamma$-ray SED. The NIR-optical jump was found to be consistent with the accretion
disk of the primary while the optical-UV could come from the little blue bump. Addition
of IC scattering of the optical-UV bump naturally reproduced the observed hardening
and the shift of the peak of the $\gamma$-ray SED \citep{2018MNRAS.473.1145K}. Recently,
the source underwent the highest ever rise in X-rays \citep{2017ATel10043....1G} and
was concurrently detected at the very high energy band (VHE, $\rm E > 100$ GeV) by
VERITAS \citep{2017arXiv170802160O} between February and March 2017 with its energy
spectrum being consistent with a power-law model of spectral index $\Gamma=-3.49$.

Here we present new observations that follow those of our study of OJ 287 in
\citet{2017MNRAS.465.4423G,2018MNRASa}
and \citet{2018MNRAS.473.1145K}, focusing on multi-wavelength (MW) spectral and
temporal variability. The paper has five sections, with \S2 describing the summary
of the compiled MW data and the reduction procedures. \S3 presents the details
of the spectral and temporal analysis with implications and insights given in \S4, followed
by our conclusions in \S5. A $\Lambda$CDM cosmology with $\rm H_0$ = 69.6 km s$^{-1}$
Mpc$^{-1}$, $\rm \Omega_M$ = 0.286 and $\Omega_\Lambda$ = 0.714 is assumed for
calculation of physical quantities. 

\section{Multi-wavelength Data and Reduction} \label{sec:data}

\subsection{Fermi $\gamma$-ray Data}

The $\gamma$-ray data belong to the Large Area Telescope (LAT), the primary instrument
on board the \emph{Fermi} space observatory. It operates primarily in scanning mode
and covers all of the sky every $\sim$ 3 hours, detecting $\gamma$-ray photons
between 20 MeV and $>$ 300 GeV \citep{2009ApJ...697.1071A}. In the present work,
we have used the latest, Pass 8\footnote{http://www.slac.stanford.edu/exp/glast/groups/canda/lat\_Performance.htm},
\citep{2013arXiv1303.3514A} instrument response processed data of the source from
31 July 2016 to 24 July 2017 (MJD: 57600 -- 57950).

The data are analyzed following the standard 
procedures\footnote{https://fermi.gsfc.nasa.gov/ssc/data/analysis/scitools/python\_tutorial.html}
using the \emph{Fermi Science Tool} version v10r0p5. For each data point, the photons
event file was generated by selecting ``source'' class events with ``evclass=128, evtype=3''
from a circular region of interest (ROI) of $15^\circ$ centered on the source with
energies between 0.1$\leq$E$\leq$300 GeV. A maximum zenith angle of 90$^\circ$ was
used to avoid the Earth's limb $\gamma$-ray photon background. The corresponding
good time intervals were generated with ``(DATA\_QUAL$>$0)\&\&(LAT\_CONFIG==1)''.
The exposure file, incorporating corrections for cuts and contribution from other
sources in the ROI and those immediately outside it, was generated on the ROI plus
an additional annulus of $10^\circ$ around it. The events were then optimized for 
the input spectral model file using the `unbinned likelihood analysis' of \textsc{Gtlike}
provided within the \textsc{Python} library of the analysis software. The input model file
was generated from the 3rd LAT catalog \citep[3FGL--gll\_psc\_v16.fit;][]{2015ApJS..218...23A}
and included the Galactic and isotropic extragalactic contributions through their
respective templates \emph{gll\_iem\_v06.fits} and \emph{iso\_P8R2\_SOURCE\_V6\_v06.txt}.

We generated the daily $\gamma$-ray light curve for OJ 287 following the above-mentioned
procedure. The source was modeled with a power-law $\gamma$-ray spectrum while the
default models, as used in the 3FGL catalog, were used for the rest of the sources.
During the optimization, the Galactic and extragalactic components were fixed to the
their best fit values obtained from the whole duration of these observations. In
addition, we extracted two $\gamma$-ray SEDs, for MJD 57600--57750 and for MJD 57750--57900.
The former corresponds to the state when the source is not detected in VHE while
the latter represents frequent detection at VHE \citep{2017arXiv170802160O}. For
this, we divided the LAT energy band into six bins, 0.1--0.3, 0.3--1, 1--3, 3--10,
10-100, and 100--300 GeV, and a power-law source spectrum was used. 
\subsection{Swift X-ray and UV/Optical Data}

{\bf X-ray:} The X-ray Telescope (XRT) on board the \emph{Swift} observatory has
a wide dynamic coverage with automatic data acquisition depending on the source
brightness \citep{2005SSRv..120..165B}. For count rates below 0.5 count s$^{-1}$, the data is recorded
in photon counting (PC) mode and this shifts to window timing (WT) mode above it for
rates up to 100 count s$^{-1}$. In the present work, we have used the pointed observations
of these two modes which are analyzed following the recommended procedures with default
parameters. The science level files were generated by processing each observation
ID via the \emph{xrtpipeline} task with latest calibration files. The
resulting events files were then analyzed within \emph{xselect} using a circular
region of $\rm 47.2''$ for OJ 287 and an annular source-free region for the background
to estimate the source flux and spectrum. The PC mode observations with source region
count rate $\gtrsim$ 0.5 count s$^{-1}$ were checked for pile-up and were corrected following the
recommended procedures\footnote{http://www.swift.ac.uk/analysis/xrt/pileup.php}
\citep[e.g.][]{2017MNRAS.472..788G}. 

The spectra were modeled in \emph{XSPEC} using ancilliary response file generated
by the \emph{xrtmkarf} task. We used a minimum of 20 counts per energy bin and the
$\chi^2$ statistics. Most of the observations were found to depart from a power-law
description at the high energy end. Thus, we additionally used a log-parabola model
based on the \emph{ftest} probability. The log-parabola description was used over
power-law whenever the \emph{ftest} probability became $\lesssim ~0.05$. During
the analysis, the neutral hydrogen column density (N$_{\rm H}$) was fixed to the
Galactic value of $\rm 2.38 \times 10^{20}~ cm^{-2}$ if the best fit resulted in
a value below, or consistent with, it. The unabsorbed flux was calculated using
the \emph{cflux} task while the X-ray SEDs were corrected for N$_{\rm H}$ absorption
using the ratio of the model without N$_{\rm H}$ (N$_{\rm H}$ = 0 without fit) and
with N$_{\rm H}$ (best fit).

\vspace*{0.05in}
\noindent
{\bf \emph{Swift}-UVOT:}
The UV-Optical observations by the Ultraviolet and Optical Telescope
\citep[UVOT;][]{2005SSRv..120...95R} on board the {\it Swift} observatory were
analysed for all the filters (v, b, u, uvw1, uvm2, uvw2). We used the available sky
images which were already corrected for any shift or rotation from the source position.
We summed the frame exposures to increase the signal to noise ratio, if any, by using
\textsc{ftool} task \emph{uvotimsum}. Low sensitivity patches on the CCD were checked
by running UVOTSOURCE on sky images of the UVOT filters using the latest small scale
sensitivity file\footnote{https://swift.gsfc.nasa.gov/analysis/uvot\_digest/sss\_check.html}.
Data points associated with the bad patches on the CCD were excluded
from the analysis. A circular region of 5\arcsec was employed for the source and
an annular region centered on the source, of inner and outer radii of 25\arcsec and
35\arcsec, respectively, was used to find the background.

\begin{figure*}
\begin{center}
\includegraphics[scale=0.95]{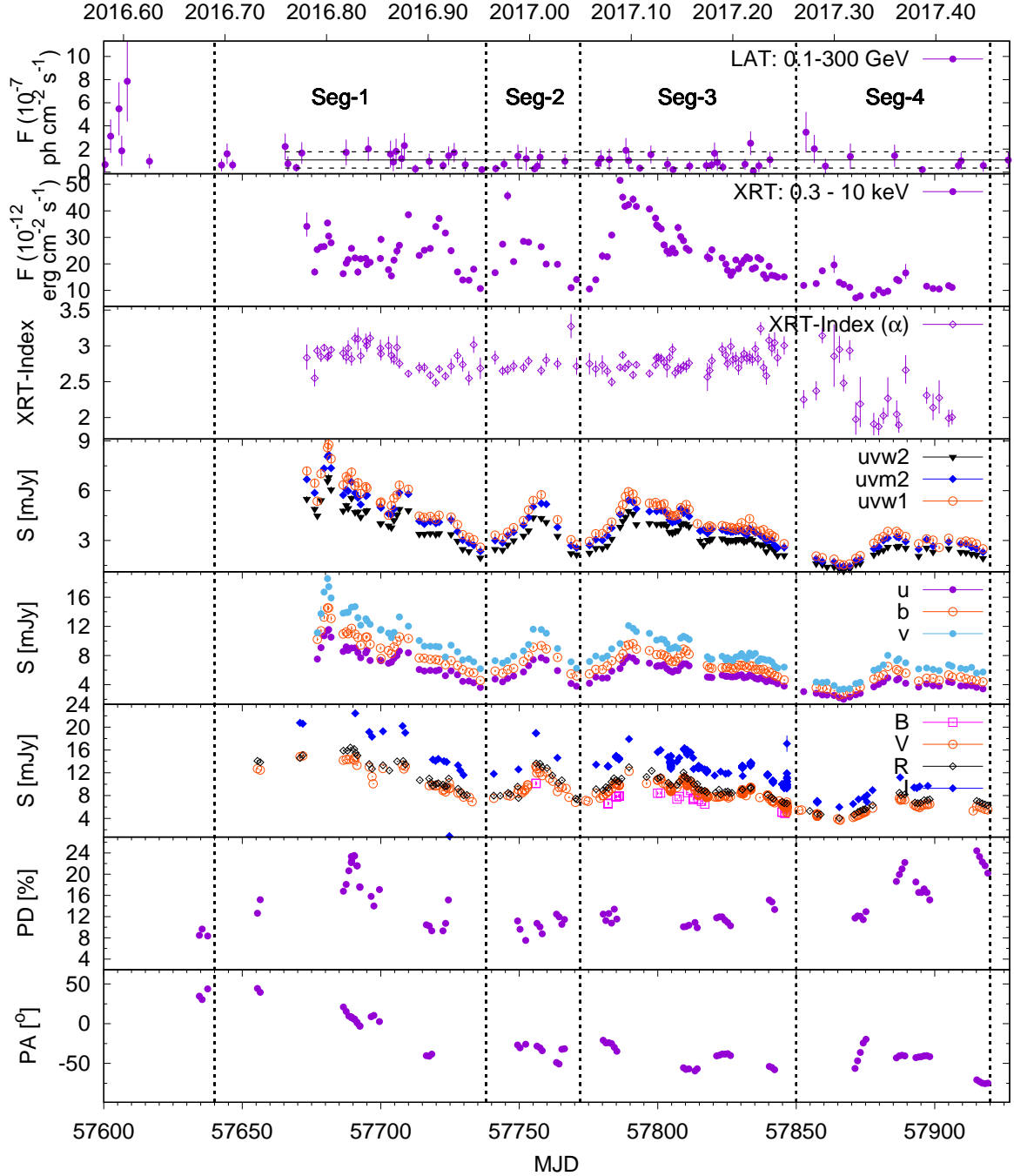}
\end{center}
\caption{Light curves of OJ 287 from $\gamma$-ray to optical bands between April
2016 -- July 2017 along with X-ray photon spectral index ($\rm\alpha$), optical ploarization angle
(PA) and degree (PD). The daily $\gamma$-ray points (LAT: 0.1 -- 300 GeV) belong to
the \emph{Fermi}-LAT observations while the X-ray (XRT: 0.3 -- 10 keV), UV (uvw2,
uvm2, uvm1) and optical (u, b, v) are from the different bands of the \emph{Swift}
observatory. The optical B, V, R, I data are from the nine ground-based 1-2 meter class
telescopes (see \S\ref{subsec:gOpt}). The vertical dashed lines delineate the four high activity
periods of the source while the horizontal solid and dashed lines in the top panel
represent the mean LAT flux and its error for the duration over which they
are drawn.}
\label{fig:mwlc}
\end{figure*}

We then extracted the flux density for the net
exposure of a particular observation by using \emph{uvotsource} with latest
calibration. The extracted light curves between MJD 57600 to MJD 57950
for every band of each observation as shown in Fig. \ref{fig:mwlc}. The fluxes have
been corrected for extinction following the description of \citet{2009ApJ...690..163R}
with an E(B$-$V) value of $\rm 0.0280 \pm 0.0008$.

\subsection{Optical Data}\label{subsec:gOpt}

{\bf Photometry and Polarimetry:} Our ground-based optical photometric data
for the blazar OJ 287 during this temporal period are taken from nine 1--2m class
optical telescopes in Japan, China, Georgia, Bulgaria (2 telescopes), Serbia, Spain,
and USA (2 telescopes). Optical photometric observations were made on $\sim$ 140
observing nights during which we collected a total of $\sim$ 1750 image frames of
OJ 287. The detailed photometric observation log  is given in
\citet{2018MNRASa}. Optical polarimetric observation of OJ 287 were taken during
$\sim$ 75 observing nights from which we obtained $\sim$ 85 polarimetric measurements. 

The data include the photometric and polarimetric observations from the public
archive\footnote{http://james.as.arizona.edu/psmith/Fermi/datause.html} of Steward
Observatory, University of Arizona, USA. The observations were carried out using
two telescopes with SPOL CCD Imaging/Spectropolarimeter instruments attached to them.
The details about the telescopes, instruments, observation and data analysis methods
are given in \citet{2009arXiv0912.3621S}.

New photometric observations for this project were carried out with seven other
optical telescopes. CCD detectors and optical BVRI broad band filters are attached
to these telescopes. The details about these telescopes and their CCDs are given
in our earlier papers \citep[e.g.][]{2014RAA....14..719H,2015MNRAS.451.3882A,
2016MNRAS.458.1127G,2018MNRASa}. Photometric data collected from these telescopes
were processed using standard image processing techniques. Image processing (i.e.
bias subtraction, flat fielding, cosmic ray removal, etc.) of different telescope
data are done using Image Reduction and Analysis Facility (IRAF)\footnote{IRAF is
distributed by the National Optical Astronomy Observatories, which are operated by
the Association of Universities for Research in Astronomy, Inc., under cooperative
agreement with the National Science Found} or locally developed procedures using
the Interactive Data Language (IDL), or using the software MaxIm DL. Instrumental
magnitudes of the blazar OJ 287 and standard stars in the blazar field are estimated
using DAOPHOT II software \citep{1987PASP...99..191S,1992ASPC...25..297S}, or MaxIm
DL. A differential photometric technique is used and the blazar OJ 287 photometric 
data is calibrated using the local standard stars \citep{1996A&AS..116..403F}. The
details of the observations and analysis procedure will be given another paper \citet{2018MNRASa}.
The extinction correction of the data is done with the E(B-V) value mentioned above,
following \citet{2011ApJ...737..103S}, while the magnitude to  flux conversion is
done using the zero point fluxes from \citet{1998A&A...333..231B}. The data from
different observatories are in excellent agreement with each other during overlapping
or nearly simultaneous measurements, and have similar measurement errors \citep[e.g.]
[]{2017MNRAS.465.4423G}.

\subsection{Multi-wavelength Light Curves}\label{sec:analysis}

The MW light curve of the source covering the broad optical-to-$\gamma$-ray portion 
of the electromagnetic spectrum, along with the X-ray power law spectral index and 
optical polarization data, is shown in Figure \ref{fig:mwlc}. The most prominent and 
interesting feature in the MW light curve is the huge flux variations in optical-to-X-ray 
bands without any significant variability at the Fermi-LAT $\gamma$-ray energies, 
though the source is  detected frequently in the Fermi-LAT bands on daily
timescales.  However, all $\gamma$-ray data are consistent with the mean and average error around
it for the duration of the MW campaign, as shown by the solid and dashed lines in fig \ref{fig:mwlc} and hence,
statistically, not considered to be variable. The optical-to-X-ray light curves, on the other hand,
show periods of high activity
which can be grouped into four segments: (MJD) 57640 -- 57738, 57738 -- 57772, 57772 --
57850, and 57850 -- 57920 based on the activity in the optical-UV light curves.
The dashed vertical lines in fig. \ref{fig:mwlc} mark these temporal durations which are
refered to hereafter as Seg-1, Seg-2, Seg-3, and Seg-4, respectively. Additionally,
we have listed the MJDs discussed/considered in the text related to the fluxes and
spectral states of the source in Table \ref{tab:impMJDs}.

\begin{table}
 \centering
 \caption{Events associated with important MJDs}
 \label{tab:impMJDs}
 \begin{tabular}{ll}
  \hline
  MJD & Event \\ \hline 
  \hline
  57552 & appearance of inverted XRT spectrum within Swift observations \\
  57786 & highest ever reported X-ray flux of OJ 287 \\
  57861 & close to minimum flux in optical-UV but flaring in X-ray \\
  57871 & minimum flux in X-ray, UV, and optical\\
  \hline
 \end{tabular}
\end{table}

The strong X-ray variability is accompanied by a change in the X-ray photon spectral
index, $\alpha$ (Power-law: $N(E) \propto E^{-\alpha}$; Log-parabola: $N(E) \propto
E^{-\alpha - \beta log(E)}$) from its normal state where a power-law index $\lesssim 2$
is typical \citep{2018MNRAS.473.1145K,2017MNRAS.468..426S,2013MNRAS.433.2380K} to
steep values (2.5 -- 3), although it appears to be reverting to its normal state
at the end of the period of these measurements. Interestingly, this change in X-ray
behavior is also coincident with the first-ever detection of the source at VHE by
VERITAS on a nightly basis \citep{2017arXiv170802160O}. This detection is also
reflected in the LAT $\gamma$-ray SED of the source, where the first half (57600--57750)
is softer compared to the latter half (57750--57900). Further, the extrapolated
VHE spectrum  during the latter half is consistent with the LAT spectrum (see Fig. \ref{fig:SED}). 

Visually, the optical-to-X-ray variability appears well correlated with simultaneous
changes within the observational cadence of different bands. However, there are
periods where  X-ray, and optical variations are at odds with each
other, especially during MJD 57850 -- 57865 when the variations at higher energies
systematically seem to precede the lower ones. The optical polarization degree (PD)
also shows significant variations and is in tune with the optical variation
\citep[e.g.][]{2018MNRASa} while the polarization angle (PA) shows a monotonic
decrement with swings ($\rm < 50^\circ$) superposed on it.

\section{Spectral and Temporal Variability Analysis and Results}

\subsection{Cross Correlation Analysis}\label{subsec:zdcf}

\begin{figure*}
\centering
 \includegraphics[scale=0.80]{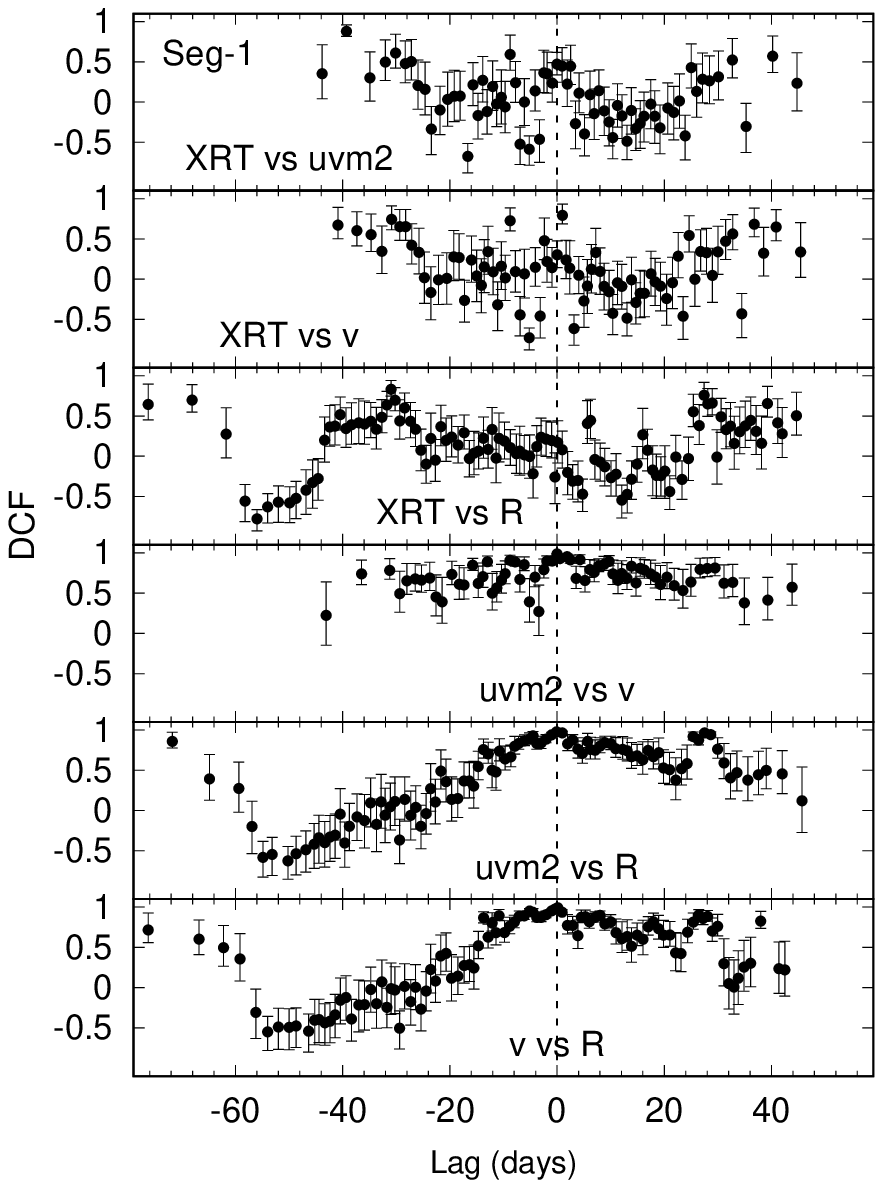}
\hspace*{0.3in} \includegraphics[scale=0.80]{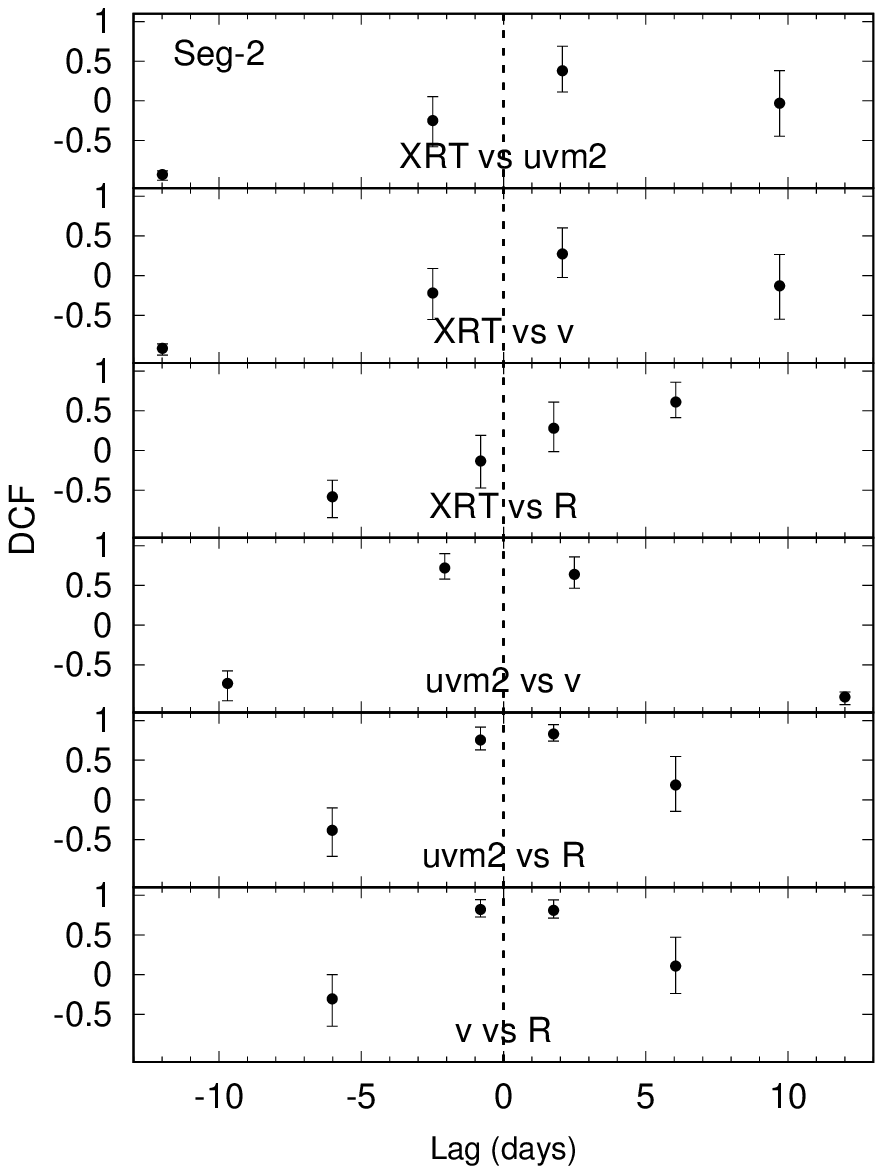}
 \includegraphics[scale=0.80]{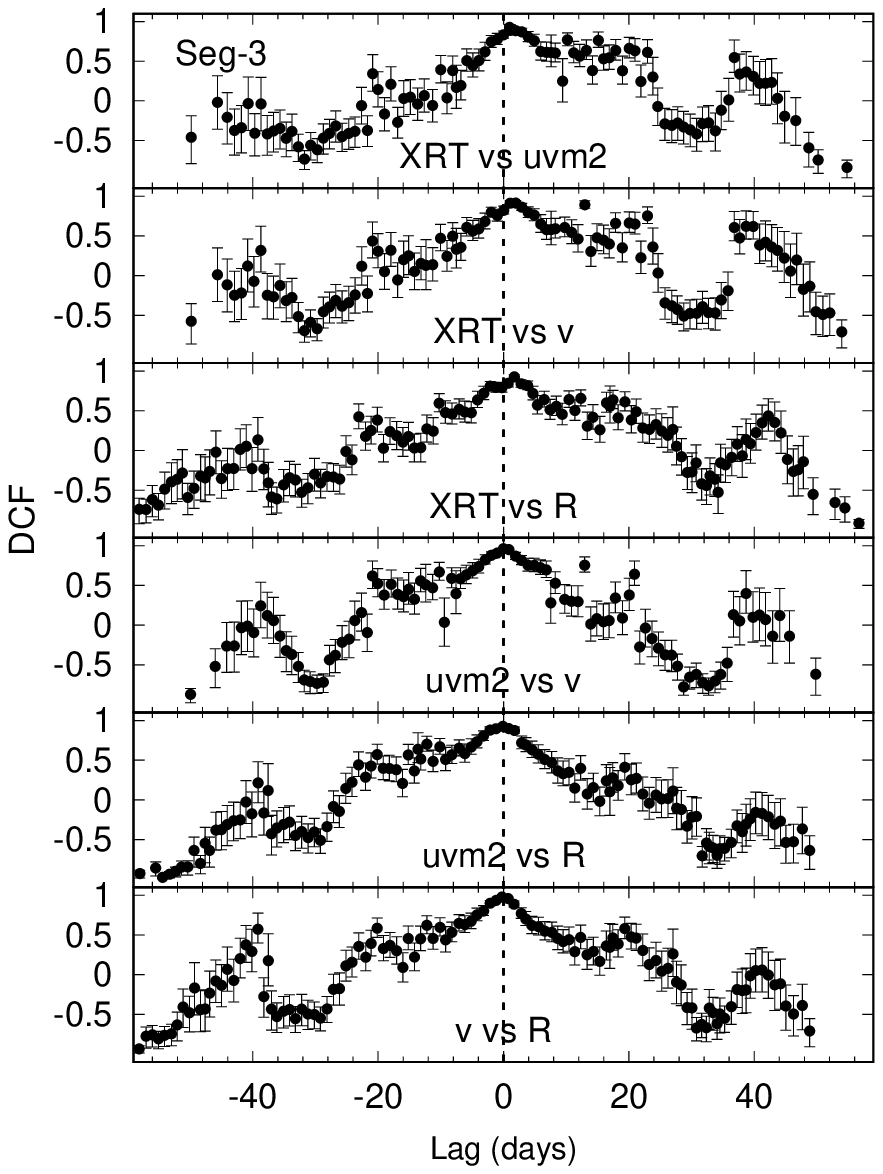}
\hspace*{0.3in} \includegraphics[scale=0.80]{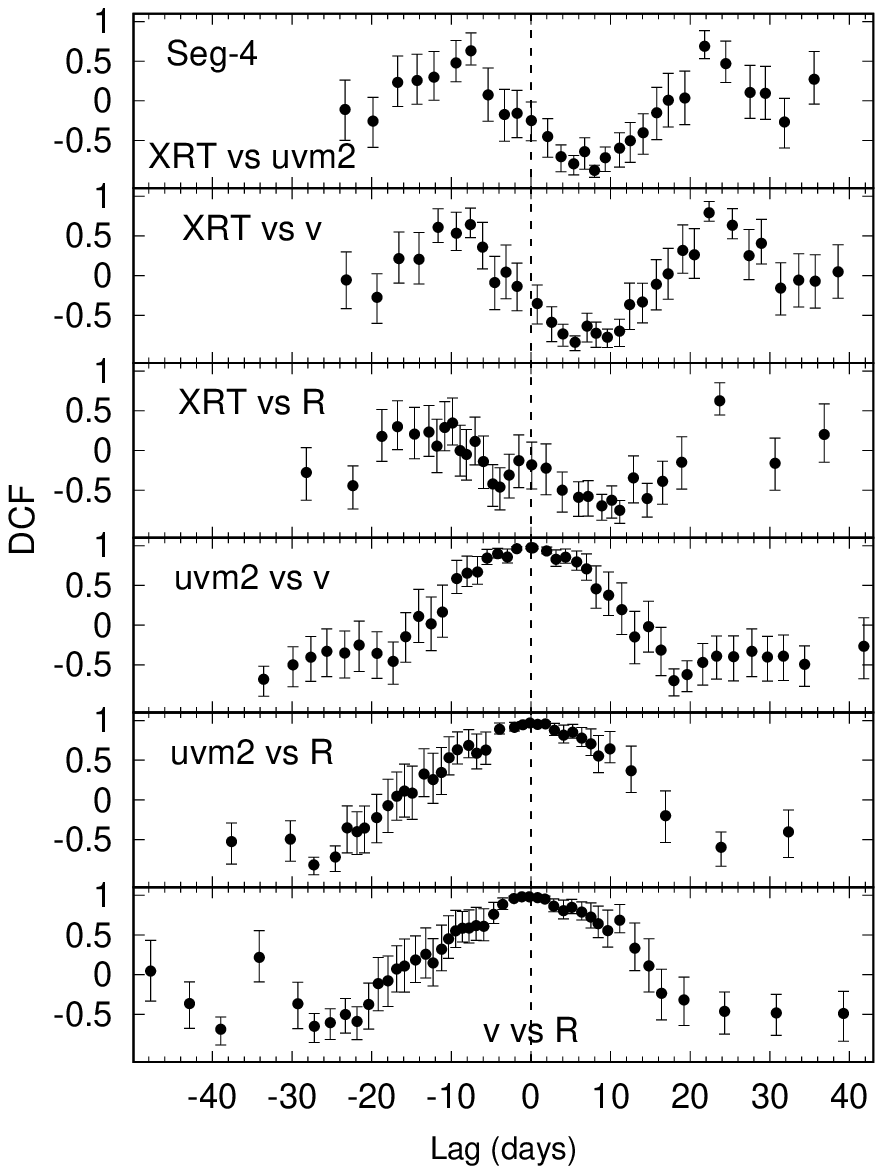}
 \caption{DCF between X-ray (XRT), UV (uvm2 band), and optical (v and R bands)
 during the four high activity periods marked in figure \ref{fig:mwlc}. The dashed
 vertical line corresponds to the zero lag. In the used DCF label ``LC1 vs LC2'',
 a positive lag means changes in LC2 lagging LC1.}
 \label{fig:zdcf}
\end{figure*}

\begin{table*}
  \centering
  \caption{Lag results for all the segments (in days)}
  \label{tab:lagResults}
  \begin{tabular}{ccccc}
  \hline
Light curves   & Seg-1 &  Seg-2   &  Seg-3 &  Seg-4 \\ \hline
XRT vs uvm2 & $-39.1^{+5.1}_{-1.4}$   & $+2.1^{+4.1}_{-2.7}$  & $+1.1^{+11.5}_{-0.5}$  & $+5.4^{+3.9}_{-1.5}$ \\ 
XRT vs v  & $+0.99^{+0.31}_{-0.46}$ & $+2.1^{+4.2}_{-3.3}$  & $+2.0^{+0.6}_{-0.9}$   & $+5.5^{+3.9}_{-1.3}$ \\
XRT vs R       & $-31.0^{+0.6}_{-0.6} $  & --                    & $+1.7^{+0.5}_{-0.5}$   & $+11.1^{+3.2}_{-5.7}$ \\
uvm2 vs v & $+0.0^{+0.4}_{-0.5}$ & $-2.1^{+5.9}_{-3.3}$  & $0.0^{+0.5}_{-0.4}$    & $+0.0^{+1.3}_{-1.5}$ \\
uvm2 vs R   & $-0.1^{+27.4}_{-0.6}$   & $+1.8^{+2.2}_{-3.1}$  & $-0.3^{+0.8}_{-0.7}$   & $-0.2^{+1.1}_{-0.8}$ \\
v vs R    & $+0.12^{+0.36}_{-0.31}$ & $-0.8^{+3.2}_{-2.4}$  & $-0.2^{+0.5}_{-0.4}$   & $-1.2^{+1.5}_{-0.7}$ \\
  \hline
  \multicolumn{5}{l}{ Seg-1: 57640-57738; Seg-2: 57738-57772; Seg-3: 57772-57850; Seg-4: 57850-57920}
  \end{tabular}
\label{tab:lagResutls}
 \end{table*}

 To understand the inter-band correlations during the four activity periods
(Segs 1--4), we performed cross-correlation  analyses between optical, UV and  X-ray
light curves\footnote{The DCF of LAT $\gamma$-ray with others result in zero/insignificant
correlation (DCF value $\lesssim$ 0.6).} using the \emph{z-transformed discreet correlation function} (zDCF) of
\citet[see also \citet{1997ASSL..218..163A}]{2013arXiv1302.1508A}. The method is
applicable to homogeneously as well as sparse, unevenly sampled data and uses
the Fisher's z-transform and equal population binning to correct the sampling
and other biases. For a given pair of light curves, it first constructs all the
possible time lag pairs using data from the two light curves. These are then sorted
according to their lags and binned by equal population to have at least 11 lag
points per bin, discarding the interdependent pairs. The correlation coefficients
are then calculated for these bins which are transformed into the z-space via
\begin{equation}
 z = \frac{1}{2} ln\left(\frac{1+r}{1-r}\right), \qquad 
 \zeta = \frac{1}{2} ln\left(\frac{1+\rho}{1-\rho}\right), \qquad
 r = tanh(z) \nonumber
 \end{equation}
where $r$ and $\rm \rho$ are, respectively, the bin correlation coefficient and the
unknown population correlation coefficient. The transformation makes the highly
skewed distribution of coefficients \emph{r} in the real domain into an approximately
 normal distribution in the z-space, thereby making the estimation of error reliable.
For an error estimate, it uses a Monte Carlo method where it constructs new light
curves by adding a random error to observed fluxes based on the measurement error
and then applies the zDCF on it. The mean and variance of these runs in the 
\emph{z}-space, when transformed  back into the real space provide the coefficients
and errors on them.

Figure \ref{fig:zdcf} shows the DCFs between possible light curve pairs for the
four high activity durations (Segs 1 -- 4) and the corresponding lag results are listed
in Table \ref{tab:lagResults}. The errors are estimated from 1000 Monte Carlo
simulations of each pair of light curves. In Figure \ref{fig:zdcf}
and Table \ref{tab:lagResults} the DCF labels tagged as ``LC1 vs LC2''  refer to
the two light curves between which the correlation is examined. In this notation,
a positive lag means the variation appears first in ``LC1'' followed by ``LC2'',
while a negative lag means the opposite.

\subsection{Spectral Energy Distributions} \label{subsec:SED}

\begin{figure*}
\centering
 \includegraphics[scale=0.65]{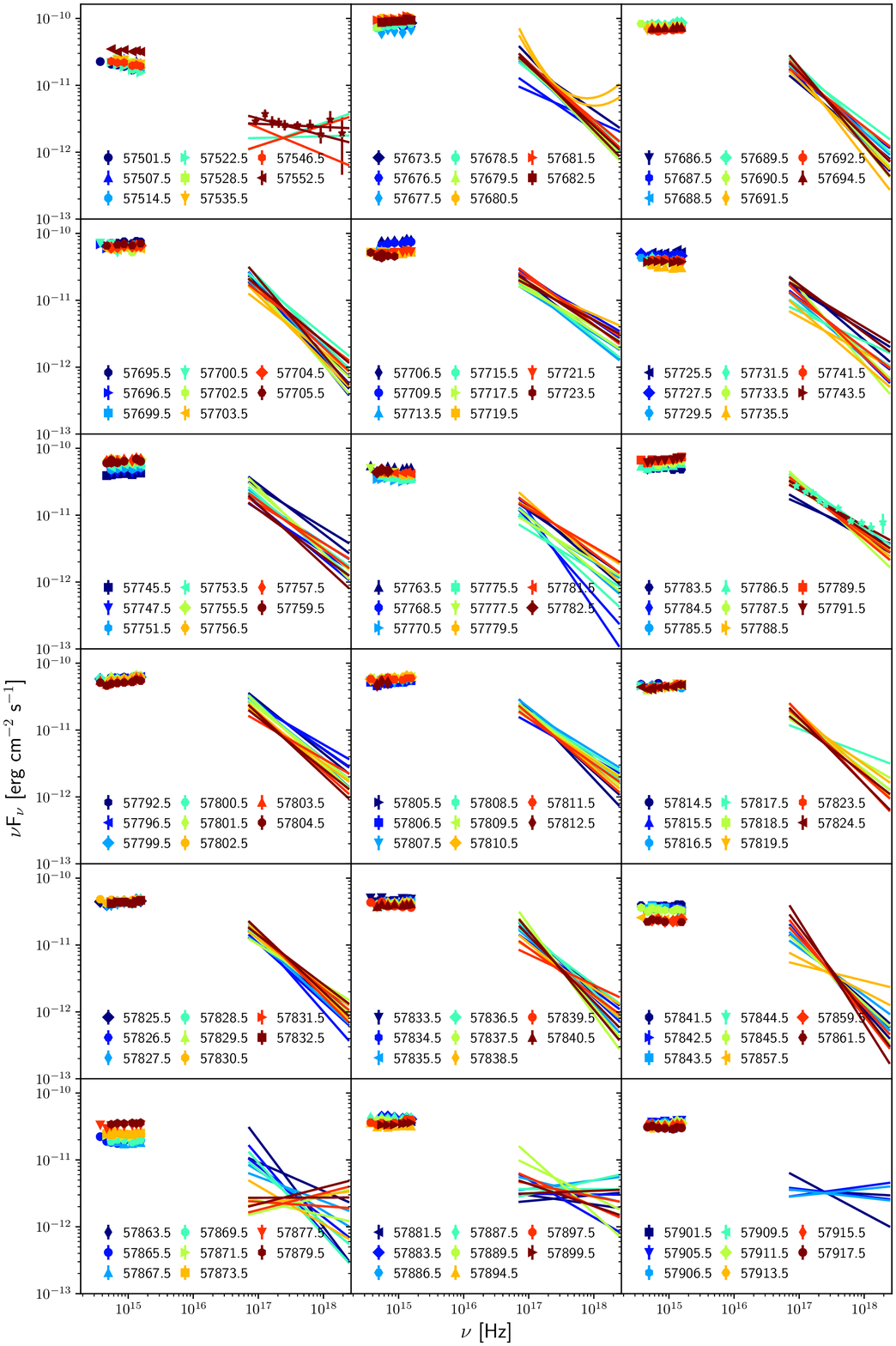}
\caption{ The daily optical to X-ray SEDs of OJ 287 between MJD 57500 -- 57920,
shown only for days having observations in at least four different bands. The optical
to UV measurements are shown by actual data points while the X-ray spectrum
is shown by the 1$\sigma$ range of the best fitted spectrum (see \S\ref{subsec:SED})
which is a good representation of the observed SED as shown by the overlaid X-ray data
in the first and the ninth plot in the respective color.}
\label{fig:irXseds}
\end{figure*}

Figure \ref{fig:irXseds} shows the  daily evolution of optical to X-ray
SEDs of OJ 287 between MJD 57500 and 57920.  Only days having data in
four different bands are shown. The X-ray SED is represented as the $1\sigma$
range around the best fit power-law/log-parabola spectrum to each of the X-ray
observation IDs  while the data belong to the optical-UV bands.
The representation provides a good overall description of the
actual SEDs except for the highest end of the spectrum for a few of the observations
as can be seen from the overlaid star data in first (for MJD 57552.5) and third
(for MJD 57786.5) rows of Figure \ref{fig:irXseds}.  

\begin{figure*}
\centering
 \includegraphics[scale=1.3]{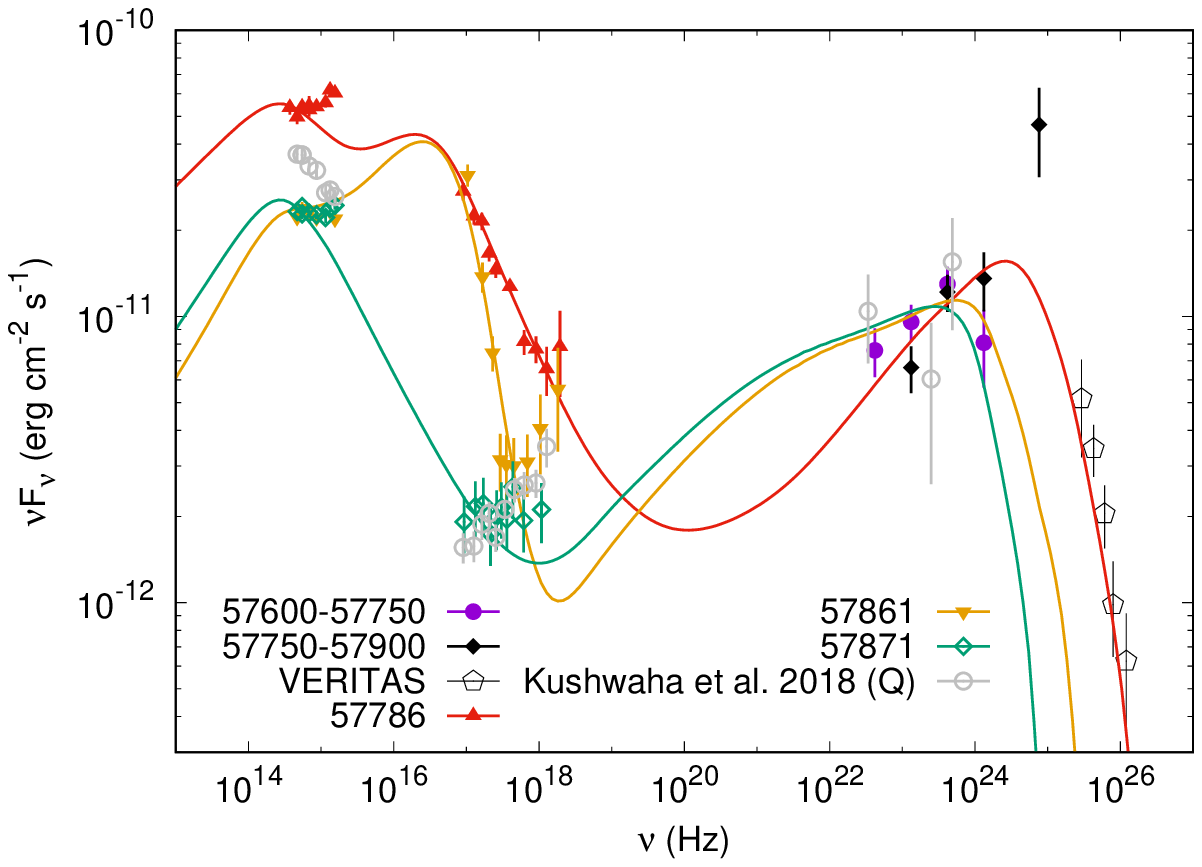}
\caption{ Overall broadband SEDs of OJ 287 at different observational epochs.
The optical-to-X-ray SEDs are for three different flux and spectral states, chosen
in such a way that optical-to-X-ray SEDs at other durations/days are in-between
these three. The optical-to-X-ray SEDs are simultaneous within a day while the
LAT spectra for these are extracted for two durations, one before (purple solid
circles) the source VHE activity as seen by VERITAS and other during it (black solid
squares). The solid curves are the total emission from a two zone model where one
zone is responsible for an LBL SED while the other for an HBL SED (see \S\ref{subsec:SED}). }
\label{fig:SED}
\end{figure*}

Particularly interesting are the strong variability (by a factor of $\sim 5$) of
the X-ray fluxes and the spectral inversion of the X-ray emission (in the sense
that the $\nu F_{\nu}$ values decline, or $\alpha > 2$) from its previously observed
normal states where they are flat or rising ($\alpha \le 2$) \citep{2018MNRAS.473.1145K,
2017MNRAS.468..426S,2013MNRAS.433.2380K,2009PASJ...61.1011S}. Within the observational
cadence, the inversion first appeared on MJD 57552 (see Fig. \ref{fig:irXseds}).
This change in X-ray spectrum is also reflected in the optical-UV SED, which also
 hardens. The spectral variations from the beginning until MJD $\sim$ 57677, show
the building up of this new spectral state, leading to an inverted spectrum. Following
MJD 57677 to 57840 the X-ray spectrum is quite stable, except for an overall change in
the normalization, which is also reflected in the optical-UV. In fact, most of
the X-ray SEDs show departures from the fitted power-law spectrum at the high
energy end of the spectrum, a typical example of which can be seen in the third
row of Fig. \ref{fig:irXseds} (see also Fig. \ref{fig:SED}), thereby revealing the
historical state that is being dominated by the new component during most of these
observations. After MJD 57860, the X-ray spectrum becomes chaotic, showing frequent
changes but with a tendency towards  its normal SED state. This is also reflected
in the optical-UV part as well and suggests the weakening of this new component.

Figure \ref{fig:SED} shows the broadband SEDs corresponding to the different flux
and spectral states of the source in optical-UV and X-rays: MJD 57871 corresponding
to the lowest optical-to-X-ray flux; MJD 57861 corresponding to a high X-ray flux but
close to the lowest in optical-to-UV; MJD 57786 corresponding to the highest X-ray flux
and a high optical-to-UV flux. The optical to X-ray SED for other flux states are in
between these three with changes mainly in the overall normalization/strength.
Since the Fermi-LAT $\gamma$-ray fluxes do not show significant variability, the
$\gamma$-ray spectrum is extracted for two equal durations: MJD 57600--57750 and
MJD 57750--57900, roughly representative of the no VHE and VHE activity reported by
 VERITAS \citep{2017arXiv170802160O}. In addition, we have plotted the quiescent
state broadband SED (Q) from \citet{2018MNRAS.473.1145K} for comparison where the
X-ray spectrum was still in its typical state. The Fermi-LAT $\gamma$-ray SEDs also
reflect the reported VHE variability, where the first half appears consistent with
the quiescent state SED clearly showing the spectral break and the shift in the peak
as reported before in \citet[see also \citet{2017arXiv170802160O}] {2018MNRAS.473.1145K}
but with much better statistics. The latter half is consistent with the reported
VHE spectrum upon extrapolation to {\it Fermi}-LAT energies.

 Observationally, the low state broadband SED (MJD 57871) is similar to
 the quiescent SED of the preceding segment \citep{2018MNRAS.473.1145K}
and thus, can be attributed to  the emission from a single zone. The SEDs
for high activity periods, on the other hand, can be characterized as a sum of an LBL SED (typical OJ
287 SED with modified $\gamma$-ray spectrum) and an HBL SED where in the latter, the
synchrotron peaks occur in the UV to soft-X-ray regions. Thus, the emission during
high activity periods is attributed to two emission zones (see arguments in \S\ref{sec:discussion})
where the first zone represents the typical source SED as presented in
\citet{2018MNRAS.473.1145K} while the second zone is responsible for an HBL-like
SED. For modelling, we have followed the approach adopted in \citet{2013MNRAS.433.2380K}
of inferring the parameters from observations directly. However, such an unambiguous
estimation of parameters is not possible in the present case due  to the strong dominance
of one  of the zones in some part of the optical to X-ray region while the other is more important in the rest (zone
one mainly in the optical, while second contributes more in UV-X-ray), thereby making the level of
flux and the spectral indices uncertain except for the low state SED (MJD 57771).
However, even in the low  state SED case, information from NIR bands (J and K) is missing
where the break due to accretion disk was seen during the preceding duration
\citep{2018MNRAS.473.1145K}. This further introduces uncertainty about the level
of thermal/non-thermal contribution \citep[see][]{2018MNRAS.473.1145K}. Within these
limits, we provide a brief description of  the model below.

For the low state, the emission region is  from a spherical blob of plasma,
moving with a bulk Lorentz factor ($\rm \Gamma$) at an angle theta ($\rm \theta$)
to the observer's line of sight. The blob is filled with a randomly oriented magnetic
field and an isotropic non-thermal electron population with a broken power-law
distribution given by (primed quantities refers to the blob frame while unprimed
to the observer frame) 
\begin{equation*}\label{eq:partdist}  
  N'(\gamma')d \gamma'=\left\{
  \begin{array}{l l}
    K \gamma'^{-p} d \gamma' & \quad ;\gamma'_{min}< \gamma'< \gamma'_b \\
    K \gamma'^{q-p} \gamma'^{-q} d \gamma' & \quad ;\gamma'_b< \gamma'< \gamma'_{max}\\
  \end{array} \right.
\end{equation*}
where $\rm p$ and $\rm q$ are respectively the spectral indices responsible for the shape of the
SED, before and after its peak. $K$ is the particle normalization
while $\gamma'$ represents the particle energy in units of rest mass energy
with $\gamma'_b$ being the break in the particle spectrum and observationally
associated with the peak of the broadband SED and $\gamma'_{min}$ and $\gamma'_{max}$
define the span of the particle spectrum. The second zone is similar to first in 
terms of constituents and parameter except for their values which correspond to
the new component with synchrotron peak in UV -- X-ray region.

The size of the first emission zone, representing a typical OJ 287 SED is $\rm 3\times10^{16} cm$,
taken from the previous study \citep{2018MNRAS.473.1145K} which corresponds to
a variability time of $\sim$ 1.5 days for a bulk Doppler factor of 10. For the second
zone, the daily VHE light curve presented in \citet{2017arXiv170802160O} show a
rise time of 2 days, while the X-ray light curve having observations on adjacent
days indicate 1 day, thereby suggesting a variability time similar to the first zone.
Thus, we have assumed the size of the second zone to be same as that of first zone,
i.e., $\rm 3\times10^{16} cm$. The jet angle to our line of sight is taken to
be $\rm 3^\circ$ while $\gamma'_{min}$ is assumed to be equal to the bulk Lorentz
factor \citep{2013MNRAS.433.2380K}.

We first start by modeling the low state SED with particle spectral indices from
the LAT $\gamma$-ray spectrum. Under the Dirac delta function
approximation of IC scattering in the Thomson regime, the observed flux before the 
SED peak due to the EC process is \citep{2018MNRAS.473.1145K,2013MNRAS.433.2380K}
\begin{equation}\label{eq:ec}
 F(\nu_{\gamma}) \propto [\Gamma(1+cos\theta)]^{(p+1)/2} [\nu_\star \delta]^{(p+5)/2} K \nu_\gamma^{-(p-1)/2},
\end{equation}
where $\rm\delta~(=[\Gamma(1-cos\theta)]^{-1})$, and $\nu_\star$ are respectively,
the Doppler factor and the photon field for the IC scattering (little blue bump/BLR:
$\rm 2.47\times10^{15}$ Hz). Similarly, the synchrotron flux after the SED peak
in the optical-NIR region,  in the Dirac delta function approximation is
\begin{equation}\label{eq:syn}
 F(\nu) \propto \delta^{(q+5)/2} B^{(p+1)/2} K \gamma_b^{(q-p)} \nu^{-(q-1)/2},
\end{equation}
where $B$ is the magnetic field in the emission region and the SED peak of the synchrotron
emission is related to Larmor frequency ($\rm \nu_L$) by
\begin{equation}\label{eq:synPeak}
 \nu_p^{syn} \approx \frac{\delta}{(1+z)} \gamma_b^2 \nu_L.
\end{equation}

Thus, for a given Doppler factor and the observed $\gamma$-ray flux, the particle
normalization can be estimated from equation \ref{eq:ec}. With this, the magnetic
filed can be estimated using equations \ref{eq:synPeak} and \ref{eq:syn} by the 
constraint that the peak is below the lowest frequency of the optical bands while at
the same time it reproduces the X-ray emission. Thus, one can iterate until the
the condition is satisfied. It should be noted that for IC of BLR, the LAT spectrum
lies in the Klein-Nishina regime \citep{2001ApJ...561..111G} and hence, slightly
harder particle spectral indices will be needed.

\begin{table*}
\centering
\caption{ SED parameters}
\begin{tabular}{l c c c}
\hline
Parameters & 57786 & 57861 & 57871  \\
\hline
Particle index before break (p)  & 2.5 (2.0) & 2.2 (2.0) & 2.2 \\
Particle index after break (q) & 4.4 (4.2)  & 4.4 (6.6) & 4.0 \\
Magnetic field (Gauss) & 11.5 (0.8) & 1.8 (1.7) & 2.5 \\
Particle break energy $(\gamma_b^\ast)$ & 1145 (36086) & 3729 (35512)  & 2537 \\
Relativistic particle energy density (erg cm$^{-3}$)& $\rm7.5\times10^{-3}$ ($\rm1.8\times10^{-3}$) & $\rm3.4\times10^{-2}$ ($\rm5.9\times10^{-4}$) & $\rm2.3\times10^{-2}$\\ 
Doppler factor & 9.3 (17.3) & 7.5 (17.3) & 8.5 \\
Jet power (logscale, erg/s) & 46.1 (45.5) & 45.7 (45.9) & 45.6 \\
Minimum electron Lorentz factor$^\ast$ & 5 (30) & 4 (30) & 5 \\
\hline
\multicolumn{3}{l}{Size of the emission region: $\rm 3 \times 10^{16}$ cm} \\
\multicolumn{3}{l}{Jet angle to the line of sight: $\rm 3^\circ$} \\
\multicolumn{3}{l}{Maximum electron Lorentz factor: $8\times 10^{6}$} \\
\multicolumn{3}{l}{$^\ast$in units of electron rest mass energy} \\
\multicolumn{3}{l}{Values in parentheses correspond to the second zone (see \S\ref{sec:discussion})} \\
\end{tabular}
\label{tab:parSEDs}
\end{table*}

Similarly, for the broadband SEDs from the duration of high activity, the particle
spectral indices can be retrieved from the corresponding X-ray and LAT/VHE spectrum
for second zone but these are uncertain for the first zone. Also, the level of optical
and $\gamma$-ray emission from the first zone is uncertain. This, thereby, opens
a huge parameter space with varied contribution from each such that the combined output
equals the observed emission level. Following the above mentioned approach, we list a
parameter set in Table \ref{tab:parSEDs} which reproduces the observed flux reasonably well,
as shown in Figure \ref{fig:SED}. 
For the optical to X-ray SED of MJD 57861, the LAT SED (57600 -- 57750) represent
the $\gamma$-ray emission as no VHE was detected after MJD 57845 \citep{2017arXiv170802160O}
while the LAT (57750 -- 57900) and VERITAS VHE spectrum represent the $\gamma$-ray
emission for the optical to X-ray SED of MJD 57786.
Finally, it should be noted that any change in the inputs used to derived the parameters
will affect the reported values. Such a systematic and extensive investigation is beyond
the scope of present work.

\section{Discussion}\label{sec:discussion}
OJ 287 has been active at optical, UV and X-ray energies since 15 November 2015,
as expected in the binary SMBH scenario for the disk impact outburst which was
observed on 5 December 2015 \citep[and references therein]{2017Galax...5...83V,
2016ApJ...819L..37V}. Coordinated observations by many observatories were made
focusing mainly on the timing and polarization in the context of the binary SMBH
model \citep{2017Galax...5...83V,2016ApJ...819L..37V} and on the optical-NIR
variability \citep{2017MNRAS.465.4423G,2017ApJ...835..275R}.  However, high activity
was also seen in X-rays and $\gamma$-rays as well, and a systematic MW investigation
by \citet{2018MNRAS.473.1145K} revealed many new features: a bump in NIR-optical
and another in optical-UV along with a hardening of the LAT $\gamma$-ray SED. Such
features have never been seen in this or any other BL Lac source to date to the
best of our knowledge. The period considered here, 31 July 2016 through
16 July 2017 (MJDs: 57600-57950) is a continuation of the work presented in
\citet{2017MNRAS.465.4423G} and \citet{2018MNRAS.473.1145K} with focus on MW aspects.
The details of optical variation on various timescales, including intra-day, along
with polarimetric variability will be presented in  the accompanying paper \citep{2018MNRASa}.

OJ 287 is continuing its surprising trends across the MW electromagnetic spectrum,
both spectrally and temporally, with many first-ever features reported for this
source recently. These include the highest-ever observed X-ray state till date
\citep{2017ATel10043....1G} and the very first detection of the source at VHE energies
\citep{2017arXiv170802160O}.  The period shows intense  variability
in X-ray to optical bands which can be grouped into four segments
(Seg: 1 -- 4) of high activity. However, the daily LAT $\gamma$-ray
light curve is statistically invariable
(see Fig. \ref{fig:mwlc}) with all data points being consistent with the band formed
by the mean flux-error around the mean flux for the considered duration. Further, most
of the photometric variations are associated with significant changes
in PD and PA  on similar timescales. The former shows a positive bias towards the
optical flux, more often displaying an increased PD with an increment in optical
flux; however, this is not a strict correlation and so does not directly reflect
the change of the photometric amplitude. The latter shows PA swings superposed on
a monotonic declining trend, often reflective of the change in the PD. 

 Cross-correlations between optical-to-X-ray bands during the four activity periods
(Segments 1 -- 4) show that the optical and UV variations are simultaneous for all of them while
the X-ray--optical/UV variations are simultaneous only for  Seg-2 and Seg-3.
During Seg-1, the results are ambiguous. It suggests optical (R) and UV(uvm2)
leading the X-rays, but at the same time also suggests a nominal lead for X-ray
with respect to optical (v) (ref Table \ref{tab:lagResults}). However, as mentioned, the optical
and UV variations are simultaneous throughout, thereby making these results suspicious.
Further, the  peak DCF values are low ($\sim$ 0.6) and are at the edge of the explored
lag duration.  During Seg-4, the X-ray leads the optical-UV emission by $\sim$
5--6 days  with variations being anti-correlated: high flux (flaring profile)
in X-ray while minima in UV/optical. This correlation is the very first
 reported for this source on short timescales (during flares). It should be
 further noted that  Seg-4 suggests some peculiar features: a systematic
trend  in observed variations, appearing
first at higher energies followed by the lower ones, e.g., a possible increment in LAT flux around
MJD $\sim$ 57855, followed by X-ray around 57810 and then finally in UV-optical around
57835. Interestingly, the fluxes at lower energies have been close to the minima of
the respective bands during these rises   seen at the higher energies.

The spectral domain also reflects the variations seen in the temporal domain and
the evolution of the optical to X-ray SEDs presented here is unique, with unprecedented
coverage and cadence. The X-ray spectra ($\rm\nu F_\nu$) are inverted throughout
these observations, as opposed to the normally rising spectrum \citep[e.g.][]
{2018MNRAS.473.1145K,2013MNRAS.433.2380K,1997PASJ...49..631I}. The effects of changes
in the X-ray emission are also seen in the optical-UV SEDs, through the concurrent
hardening/softening and/or change in level of emission (Fig. \ref{fig:irXseds}).
The actual X-ray SED data overlaid in the third row of Figure \ref{fig:irXseds}
(see also Fig. \ref{fig:SED}) show departures from the best-fitting power-law at
the highest end of the X-ray spectrum, thereby revealing the nature of the emission
feature. Comparison with a quiescent state SED in the MW study of the preceding
period as presented in \citet{2018MNRAS.473.1145K} suggests that the inverted spectrum
is a new emission component, overshadowing the typical X-ray SED of the source. Such
inverted as well as flat X-ray spectra have been seen before for OJ 287 \citep[and
references therein]{2017MNRAS.468..426S,1997PASJ...49..631I}, but they lacked the
MW coverage and cadence that we now have. Finally, we note that OJ 287 seems to
be returning towards its typical X-ray emission spectrum by the end of these
observations (MJD $\sim$ 57870) where the variation becomes chaotic, strongly
suggesting we are seeing the weakening of the new component. This can be seen
clearly in the  quiescent SED corresponding to the MJD
57871 in figure \ref{fig:SED}  vis-a-vis the quiescent SED in the preceding
phase \citep{2018MNRAS.473.1145K}.

The uniqueness of the spectral evolution as plotted in Figure \ref{fig:irXseds}
allowed us to pinpoint the instance of transition from its historical well-known state
\citep{2018MNRAS.473.1145K,2017MNRAS.468..426S,2013MNRAS.433.2380K,2009PASJ...61.1011S}
to the present inverted one at around  MJD 57752. On the other
hand, comparison with flux and spectral state during the previous study by
\citet{2018MNRAS.473.1145K} allows us to track the evolution to this state. This
comparison suggests that the build up of the new component had already started
around MJD 57500  (see Fig. \ref{fig:irXseds}) when the X-ray SED
started becoming flat. However, this may not be the true signature of a new component
since similar flat X-ray SEDs have been inferred before by \citet{2017MNRAS.468..426S}
without any subsequent enhancement in X-ray flux variability. Further, the subsequent
gap in X-ray observations does not allow an unambiguous claim that it is the sign
of a new component. Nonetheless, the observed spectral variation, along with the
temporal flux and the concurrent VHE detection on a nightly basis, suggesting significant
variation with X-rays \citep{2017arXiv170802160O}, clearly show that the bulk of
these variations result from the new component.

The rapid variation in the PD, the power-law broadband nature of the new X-ray
component extending down to optical energies, and the concurrent VHE detection
suggests the new component to be non-thermal, with the emission originating within
the jet.    Observationally, the overall broadband SEDs (see Fig.
\ref{fig:SED}) during a high activity period appears to be a combination of typical
OJ 287 SED with a modified $\gamma$-ray
spectrum (as seen during the preceding phase \citep{2018MNRAS.473.1145K} and an
HBL SED. In the latter, the first peak in broadband SED occurs in UV-soft-X-ray
region while the second peaks in the $\gamma$-rays, with emission extending to VHEs.
 Further, the strong correlation between the X-ray and the VHE emission reported in
\citet{2017arXiv170802160O} imply co-spatial emission and favors a leptonic origin.

Investigation of MW emission in OJ 287 during the previous adjacent period by
\cite{2018MNRAS.473.1145K} shows BLR emission (blue bump) and a strong accretion
disc emission  \citep[see also][]{2010A&A...516A..60N}. Thus, for VHE photons to be observed, the emission region has to be
beyond the BLR region to avoid pair absorption \citep[e.g.][]{2014MNRAS.442..131K}.
 Further, the strong variability in optical to X-ray bands while no significant variability is seen
in the LAT $\gamma$-ray band, along with X-ray spectra showing a rising trend at the
high energy end at the level of the quiescent X-ray spectrum  during the preceding period, suggest
the observed strong variability comes from another region. Thus, the simplest description
for the broadband SEDs during the high activity period would involve two independent emission
regions: one at the sub-pc scale as argued in
\citet{2018MNRAS.473.1145K}, responsible for the general OJ 287 spectrum, and
another at the parsec scale for the VHE emission. Similar suggestions have been
made  based on X-ray emission during earlier observations \citep{1997PASJ...49..631I,
2001PASJ...53...79I,2009PASJ...61.1011S}. Other leptonic emission models involving
two or multiple zones where there  are tighter connections between the zones are
also possible, but then for VHE to be observed, this has to be located at or beyond
the boundary of the BLR. In our model here, the first zone represents the typical
emission of OJ 287 with $\gamma$-rays mainly arising as a result of IC of BLR component
as in \citet{2018MNRAS.473.1145K}. The second zone is located beyond BLR and thus,
only SSC and IC of infrared photons of around 250 K, as found by \citet{2013MNRAS.433.2380K}
are contributing to $\gamma$-ray  and VHE energies.

Since the observed emission is possibly the sum of contributions from two regions,
a varied level of contribution from each with the final sum fixed to observed values,
is plausible. Hence, many parameter sets may be possible and a systematic study
with information in IR and VHE band is needed to remove degeneracies. Here, in Table \ref{tab:parSEDs},
we list a  parameter set which represents the observations reasonably, with
the second zone parameter value given in parentheses  (see \S\ref{subsec:SED}
for details). The corresponding model flux is plotted in Figure \ref{fig:SED}.
In this model, the VHE spectrum is considered only for the optical to X-ray SED of
MJD  57786, emanating from zone 2, which is responsible for the
HBL like spectrum. The low state SED corresponding to MJD 57871 is modeled assuming
a single zone emission as it is similar to the quiescent state SED of the preceding
period (gray data, fig \ref{fig:SED}) except with a lower optical flux.

Being a potential  binary SMBH system, OJ 287 presents a very complex dynamical
system during the close encounter periods, particularly considering that each SMBH
may have its own magnetized jet and accretion disc. The complexity of the interaction
can be understood from the behavior of the high energy emission during the claimed
impact encounters \citep{2016ApJ...819L..37V}. Barring the current epoch, the source
has never been detected at VHE energies \citep[and references therein]{2009PASJ...61.1011S,
1997PASJ...49..631I}. However, an inverted X-ray spectrum, such as seen in the present work, seems
to have been present around the 1983 impact \citep[and references therein]
{2001PASJ...53...79I,1997PASJ...49..631I} while the 1994 spectrum was a mixture of
both, declining below 2 keV and rising above it. The more recent impact events in  2005
and 2008, however, had the typical (uprising) X-ray spectrum. The
X-ray variability has also been very different during the impact events. The December
5, 2015 flare showed a factor of $\sim$ 2 brightening \citep{2018MNRAS.473.1145K}.
A similar X-ray brightening was seen in the November 2008 impact with a significant
detection in hard X-rays as well \citep{2009PASJ...61.1011S}, while no elevated X-ray
emission was seen during the 1994 impact event \citep{1997PASJ...49..631I}.

The dynamics of close encounters in the binary SMBHs model of OJ 287 \citep{1996ApJ...460..207L}
has been explored by \citet{2013ApJ...764....5P} in connection with observations
via hydrodynamical modeling of the accretion disk  with particles interacting
hydrodynamically. As per the details in \citet{1996ApJ...460..207L}, the latest disk
impact occurred on 2013.45 (MJD 56457.25),  thereby releasing a hot bubble of gas. As a
result, it predicted the beginning of an optical/UV flare on 2015.96 $\pm$ 0.12
(MJD 57373 $\pm$ 43.8).  The rise was in fact seen on MJD 57351 
\citep{2016ApJ...819L..37V}, peaking on MJD 57361. Interestingly, the signature
of the accretion disc in the OJ 287 was  observationally seen around the
 time of the predicted disk impact (MJD 56439) by \citet{2018MNRAS.473.1145K}.
Further investigation by \citet{2013ApJ...764....5P} suggests escape of particles vertically from the 
disc starting in  2014.2 (MJD 56731) and reaching its peak around 2014.6 (MJD 56877),
then falling to minimum on  2015.1 (MJD 57059.5), followed by a second peak on
2015.3 (MJD 57132.5) \citep{2017Galax...5...83V}. Thus,
if the current activities are related to these particles reaching the blazar zone
where they can further accelerate to ultra-relativistic velocities in shocks
\citep{2016Galax...4...22B,1985ApJ...298..114M} or magnetic reconnection sites
\citep[e.g.][]{2016ApJ...824...48S}, the location of the blazar zone can be
estimated. Associating the highest X-ray flux on MJD 57786 ($\sim$ 2017.09)
with the escape peak of 2014.6  (MJD 56877), the observed time difference $\rm \Delta T_o \sim 1.5
$ years transforms to $\sim$ 11.5 years ($\rm\Delta T_o \delta/(1+z)$) or $\sim$ 3.5
parsec in the jet frame, assuming a constant Doppler factor ($\delta$) of $\sim$ 10,
consistent with a parsec scale origin, as argued in previous works \citep{2013MNRAS.433.2380K,
2012ApJ...747...63A} and the current VHE detection. Though previous observations
lacked a good polarimetric cadence, the detection of strong changes in polarization
properties showing both systematic and chaotic features during the ongoing activity
\citep{2018MNRASa}, along with the concurrent detection of high energy emission,
clearly suggest a dynamic role of the magnetic field associated with shock and
turbulence region \citep{2016Galax...4...22B}. In fact, the fractional polarization
changes from a systematic variation to a chaotic one during the VHE detection duration
and then returns again to a systematic one at the end \citep{2018MNRASa}, 
suggesting an important role for turbulence during the VHE activity.
Thus, to gain better insights, a self-consistent account of the relativistic SMBH
evolution considering the effects of magnetic fields is needed, though this would
be far from simple \citep{2017Galax...5...83V}.

\section{Conclusions}\label{sec:conclude}
We systematically investigated the spectral and temporal features of MW emission
from OJ 287 during April 2016--July 2017 period.   The period corresponds
to the historic X-ray variability and a concurrent, very first, detection of the
source at the VHE energies by VERITAS. 

The optical to X-ray emissions show intense  variability while no appreciable variability
is seen in the LAT $\gamma$-ray light curve. Cross-correlation analyses between
optical, UV, and X-ray during four continuous segments of the light curves resulted
a simultaneous variations at optical-UV energies throughout. The optical/UV-X-ray
correlation, on the other hand, is simultaneous only during the two middle segments. 
In the  first segment the  X-ray may lag the UV/optical, while it leads them by
$\sim$ 5-6 days during the last segment with the X-ray being essentially anti-correlated
with respect to the optical/UV.  Further, the last segment also
shows a hint of systematic variation with changes first appearing at the higher
energies followed by the lower energies. Most of the X-ray spectra show departure from a
power-law description at the highest end, which is  consistent with a log-parabolic
spectrum with a rising trend in $\nu F_\nu$ plots.  The departure
in the X-ray spectrum is consistent with the level of quiescent X-ray emission
found in \citet{2018MNRAS.473.1145K}. The combination of these results
suggests that during this period most of  the X-ray and VHE photons arise from a
different region than do the lower-energy emission. The new component looks typical
of HBL SEDs, having the first SED peak in UV-soft-X-ray energies with emission
extending to  VHEs. The overall SEDs during high activity episode are
sum of the normal OJ 287 SED (LBL) and HBL SED, and can be reproduced in a
two-zone scenario with the second zone located at parsec scales beyond BLR
region being responsible for the new X-ray component and the VHE emission.
 Most of the variations are associated with strong changes in PD as
well as in PA. In addition, the duration of VHE detection corresponds to a change
of fractional polarization from systematic and chaotic \citep{2018MNRASa},
suggesting that the current historic high energy activity is associated with
the dynamics of the magnetic field and turbulence.

\section*{Acknowledgements}
This research has made use of data, software and web tools of High Energy Astrophysics 
Science Archive Research Center (HEASARC), maintained by NASA's Goddard Space Flight 
Center. Observations obtained by the Fermi/Steward Observatory blazar monitoring
program have been funded through NASA/Fermi Guest Investigator grants NNX08AW56G,
NNX09AU10G, NNX12AO93G, and NNX15AU81G.

PK acknowledges support from FAPESP grant no.\ 2015/13933-0. ACG is partially supported by 
Indo-Poland project No. DST/INT/POL/P–19/2016 funded by Department of Science and Technology 
(DST), Government of India, and also partially by the Chinese Academy of Sciences (CAS) President's 
International Fellowship Initiative (PIFI) grant no.\ 2016VMB073. PJW is grateful for hospitality 
at KIPAC, Stanford University, and SHAO during a sabbatical. HG is sponsored by a CAS Visiting 
Fellowship for researchers from  developing countries, CAS PIFI (grant no.\ 2014FFJB0005), 
supported by the NSFC Research Fund for International Young Scientists (grant no.\ 11450110398) 
and also supported by a Special Financial Grant from the China Postdoctoral Science Foundation 
(grant no.\ 2016T90393). EMGDP acknowledges support from the Brazilian funding
agencies FAPESP (grant 2013/10559-5) and CNPq (grant 306598/2009-4). The Abastumani team acknowledges 
financial support by the Shota Rustaveli National Science Foundation under contract FR/217950/16. 
OMK also acknowledges grants NSFC11733001 and NSFCU1531245 of the China NSF. The work of the Bulgarian 
team was partially supported by the Bulgarian National Science Fund of the Ministry of Education and 
Science under grants DN 08-1/2016 and DN 18-13/2017. GD and OV gratefully acknowledge 
the observing grant support from the Institute of Astronomy 
and Rozhen National Astronomical Observatory, Bulgaria Academy of Sciences, via bilateral 
joint research project ``Study of ICRF radio-sources and fast variable astronomical objects". 
This work is a part of the project nos.\ 176011, 176004, and 176021 supported by the Ministry 
of Education, Science and Technological Development of the Republic of Serbia
SMH's work is supported by the National Natural Science Foundation of China under grants nos.\ 11203016 
and 11143012, and also partly supported by Young Scholars Program of Shandong University, Weihai, 
China. MFG is supported by the National Science Foundation of China (grants 11473054 and U1531245). 
ZZ is thankful for support from the CAS Hundred-Talented program (Y787081009).

\end{document}